\documentclass{article}
\usepackage{amsfonts,amssymb, amsmath}

\usepackage[cp1251]{inputenc}
\usepackage[english]{babel}

\newtheorem{prop}{Proposition}

\def\bq{ \begin{equation}}
\def\eq{ \end{equation}}
\def\ben{ \begin{eqnarray}}
\def\en{ \end{eqnarray}}

\begin{document}


\title{On   superintegrable systems separable in Cartesian coordinates}
\author{Yu. A. Grigoriev,  A.V. Tsiganov\\
yury.grigoryev@gmail.com, andrey.tsiganov@gmail.com\\
Saint Petersburg State University
}%
\date{}
\maketitle
\begin{abstract}
We continue the study of superintegrable systems of Thompson's type separable in Cartesian coordinates. An additional integral of motion for these systems is the polynomial in momenta of $N$-th order which is a linear function of angle variables and the polynomial in action variables. Existence of such superintegrable systems is naturally related to the famous Chebyshev theorem on binomial differentials.\\
\par\noindent
Keywords: Superintegrable systems, higher order integrals of motion, Fokas-Lagersrom system
 \end{abstract}

\section{Introduction}
\setcounter{equation}{0}
In 1984 Thompson proved superintegrability of the Hamiltonian
\[
H=p_x^2+p_y^2+a (x-y)^{-\dfrac{2}{2n-1}}\,,\qquad n\in\mathbb Z_{+}\,,
\]
where $n$ is an arbitrary positive integer  \cite{tom84}.  To simplify the notation it is best to make a 45 degree rotation $q_{1}=x+y$ and $q_2=x-y$ as in \cite{hiet87}. Such superintegrable systems are still being studied up till now, see \cite{pw17a,grav04,nieto17,pw17} and references within.

In this note we prove that dynamical system with Hamiltonian
\bq\label{ham-add}
H=p_1^2+p_2^2+ a q_1^{M_1}+bq_2^{M_2}\,,\qquad a,b\in\mathbb R\,,
\eq
is superintegrable, if exponents $M_1$ and $M_2$ belong to the following  sequence  of positive rational numbers
\bq\label{seq1}
M=0, 1,\dfrac{1}{2},\dfrac{1}{3},\dfrac{1}{4},\cdots, \dfrac{1}{n}\,,\qquad n\in\mathbb{Z}_+\,,
\eq
or sequence of negative rational numbers
\bq\label{seq2}
M=0,-2,-\dfrac{2}{3},-\dfrac{2}{5},-\dfrac{2}{7},\cdots,-\dfrac{2}{2n-1}\,.
\eq
These two sequences of exponents are distinguished according to the Chebyshev theorem on binomial differentials \cite{cb}.  The corresponding additional  first integral is a polynomial with respect to momenta $p_1$ and $p_2$.

 We also discuss  nonseparable systems with Hamiltonians
\bq\label{ham-mul}
H=p_1^2+p_2^2+\left(a q_1^{M_1}+b\right)q_2^{M_2}\,,
\eq
where $M_{1,2}$ belong to (\ref{seq1}-\ref{seq2}) and present a new integrable deformation of the  Fokas-Lagerstrom system \cite{fl80,hiet87}. The corresponding integral of motion is a polynomial in the momenta of the sixth degree.

\section{Thompson's type systems}
There are many integrable and superintegrable systems with algebraic potentials, see \cite{ran13,fl80, gram84, nieto17,hiet87, holt82, ts11,am16,pw11,tom84}.  For arbitrary rational  $M_{1,2}$   Hamiltonian  $H$ (\ref{ham-add}) is  also an algebraic function  well-defined in some part of the plane.  In the same domain of definition we introduce variables
\bq\label{x-12g}
\begin{array}{ll}
I_1=p_1^2+aq_1^{M_1}\qquad & I_2=p_2^2+bq_2^{M_2}\,,
\\ \\
\omega_1=-\displaystyle \int^{q_1} \dfrac{dx}{\sqrt{p_1^2+aq_1^{M_1}-ax^{M_1}}}\,,\qquad &\omega_2=-\displaystyle \int^{q_2} \dfrac{dx}{\sqrt{p_2^2+bq_2^{M_2}-bx^{M_2}}}\,,
\end{array}
\eq
with canonical Poisson brackets
\[
\{\omega_j,I_k\}=\delta_{jk}\,,\qquad \{I_j,I_k\}=\{\omega_j,\omega_k\}=0\,,\qquad j,k=1,2,
\]
and equations of motion
\[
\dot{I}_{1,2}=0\,,\quad \dot{\omega}_{1,2}=\dfrac{\partial H}{\partial I_{1,2}}=1\,,\quad\mbox{with}\quad H=I_1+I_2\,.
\]
For the completely integrable system the Liouville-Arnold theorem implies that almost all points of the phase space are covered by a system of open toroidal domains with the action-angle coordinates $I_1,\ldots,I_n; \omega_1,\ldots,\omega_n$. In these coordinates  the completely integrable system  has the form
\bq\label{st-eq-aa}
\dot{I}_{k}=0\,,\qquad \dot{\omega}_{k}=\dfrac{\partial H}{\partial I_{k}}\,,\qquad k=1,\ldots,n,
\eq
and  symplectic structure  is canonical $\Omega=\sum dI_k\wedge d\omega_k$ \cite{arn}.

The variables  $I_{1,2}$ and $\omega_{1,2}$ (\ref{x-12g}) satisfy standard equations of motion (\ref{st-eq-aa}) and have canonical Poisson structure  $P=\Omega^{-1}$. So, we will call them the formal action-angle variables which are well-defined functions on the original Cartesian variables  only in some part of the cotangent bundle to plane.

By definition Hamiltonian $H$ (\ref{ham-add}) is in the involution with action  variables $I_{1,2}$ and with any function on the difference of the angle variables
\[
X=F(I_1,I_2,\omega_1-\omega_2),
\]
see discussion in \cite{ts08,ts08a,ts09,ts12}.  Below we prove that $X$ is the polynomial in momenta $p_{1,2}$ if  $M_{1,2}$ belong to (\ref{seq1}) or (\ref{seq2}) because in this case $\omega_{1,2}$ are given by elementary functions.  More general case  when some function on difference $\omega_1-\omega_2$ are elementary functions on original variables we do not consider here, see discussion and examples in \cite{gon12,ts08,ts08a,ts09,ts12}.

Let us recall that expressions of the form
\[ x^m(\alpha+\beta x^n)^pdx\,,\]
where $\alpha,\beta$ are arbitrary coefficients and $m,n,p$ are rational numbers, are called differential binomials. According to the Chebyshev theorem \cite{cb} integrals on differential binomials
\[
\int x^m(\alpha+\beta x^n)^pdx\,,
\]
can be evaluated in terms of elementary functions if and only if:
 \begin{enumerate}
  \item $p$ is an integer, then we  expand  $(\alpha+\beta x^n)^p$ by the binomial  formula  in order to rewrite the integrand as a rational function of simple radicals $x^{j/k}$. Then we make a substitution $x=t^r$, where $r$ is the largest of all denominators $k$, remove the radicals entirely and obtain integral on rational function.
  \item $\dfrac{m+1}{n}$ is an integer, then we set $t=\alpha+\beta x^n$ to obtain integral
  \[
 \int x^m(\alpha +\beta x^n)^pdx=\frac{1}{2}\,\beta^{-\frac{m+1}{n}}\int t^p(t-\alpha)^{\frac{m+1}{n}-1} dt
  \]
  which belongs to Case 1.
    \item  $\dfrac{m+1}{n}+p$ is an integer, then we transform the integral by factoring out $x^n$
    \[
   \int x^m(\alpha +\beta x^n)^pdx=\int x^{m+np}(\alpha x^{-n}+\beta)^pdx\,.
    \]
    The result is a new integral of the differential binomial which belongs to Case 2.
\end{enumerate}
In our case  (\ref{x-12g})   we have
\[\alpha=I_{1,2}\,,\qquad \beta=1\,\qquad m=0, \qquad n=M,\qquad p=-1/2.\]
Hence action variables $\omega_1$ and $\omega_2$ is expressed via elementary functions only if
\[
 \dfrac{1}{M} \, \mbox{is integer}\qquad\mbox{or}\qquad   \dfrac{1}{M} -\dfrac{1}{2}\, \mbox{is integer}.
\]
In order to avoid logarithmic term $\ln(t)=\int t^{-1}$ in (\ref{x-12g}), which is also an elementary function,  we have to consider only  zero, positive and negative values of $M$, respectively.
\par\noindent
For $M_k$ from  (\ref{seq1})  action variables (\ref{x-12g}) are
\[
 M_k=0\,,\qquad \omega=\dfrac{2q_k}{p_k}\,,\qquad M_k=\dfrac{1}{n_k}>0\,,\qquad  \omega_k = \mbox{polynomial of order}\, 2n_k-1.
\]
For $M_k$ from  (\ref{seq2})  action variables (\ref{x-12g}) are
\[
 M_k=0\,,\qquad \omega=\dfrac{2q_k}{p_K}\,,\qquad M_k=-\dfrac{2}{2n-1}<0\,,\qquad  \omega=\dfrac{ \mbox{polynomial of order}\, 2n_k-1}{I_k^{n_k}},
\]
where $I_k$, $k=1,2$, is the corresponding action variable.  Let us show a few explicit formulae for positive exponents
\[
M_2=1\,,\quad \omega_2=\dfrac{p_2}{b},\qquad M_2=\dfrac{1}{3}\,,\quad \omega_2= \dfrac{p_2(3b^2q_2^{2/3}+4bq_2^{1/3}p_2^2+8/5p_2^4)}{b^3}\,,
\]
and negative exponents
\[
M_2=-\dfrac{2}{3}\,,\quad \omega_2= -\dfrac{p_2(3bq_2^{1/3}+q_2p_2^2)}{2\left(p_2^2+bq_2^{-2/3}\right)^2},
\qquad
M_2=-\dfrac{2}{5}\,,\quad\omega_2= -\dfrac{p_2(5bq_2^{1/5}+10/3 b q_2^{3/5}p_2^2+q_2p_2^4)}{2\left(p_2^2+bq_2^{-2/5}\right)^3}.
\]
Other partial or generic expressions for integrals may be found in textbooks, tables of  integrals or any computer algebra system.

\begin{prop}
A Hamiltonian system defined by $H$ (\ref{ham-add}) has a polynomial first integral $X_{N}$ of order $N$, if  $M_1$ and $M_2$ belong to (\ref{seq1}) or (\ref{seq2}):
\begin{enumerate}
  \item if $M_1=1/n_1$ and $M_2=1/n_2$, then
  \[X_{2n-1}=\omega_1-\omega_2,\qquad\mbox{where}\qquad n=\max(n_1,n_2);\]
  \item if $M_1=-2/(2n_1-1)$ and $M_2=-2/(2n_2-1)$, then
  \[X_{2n-1}=(\omega_1-\omega_2) I_1^{n_1}I_2^{n_2},\qquad\mbox{where}\qquad n=n_1+n_2;\]
  \item if $M_1=1/n_1$ and $M_2=-2/(2n_2-1)$, then
   \[X_{2n-1}=(\omega_1-\omega_2)I_2^{n_2},\qquad\mbox{where}\qquad n=n_1+n_2;\]
   \item if $M_1=0$ and $M_2=1/n$, then
  \[X_{2n}=p_1(\omega_1-\omega_2)\,,\qquad\mbox{where}\qquad p_1=\sqrt{I_1};\]
   \item if $M_1=0$ and $M_2=-2/(2n-1)$, then
\[X_{2n}=p_1(\omega_1-\omega_2)I_2^{n}\,,\qquad\mbox{where}\qquad p_1=\sqrt{I_1}.\]
\end{enumerate}
This integral of motion $X_N$ is  functionally independent from $I_{1,2}$ (\ref{x-12g}).
\end{prop}
Cases 1 and 5 were studied in \cite{nieto17}  and \cite{tom84}, respectively.

Let us show some ``compact'' examples of polynomial integrals $X_N$ with $N=8$:
\[
V=aq_2^{1/4}\,,\quad X_8=p_1\left[p_2^7+\frac72 p_2^5V +\frac{35}{8}p_2^3V^2+ \frac{35}{16}p_2V^3\right]+\frac{35a^4q_1}{128}\,,
\]
and for $V=aq_2^{-2/7}$
\[
X_8=p_2^7(p_1q_2-p_2q_1)+p_2^5\left(\frac{21}{5}q_2p_1-4q_1p_2\right)V+p_2^3\left(7q_1p_1-6q_2p_2\right)V^2+
p_2\left(7q_2p_1-4q_1p_2\right)V^3-q_1V^4\,.
\]
Here we multiply the expressions from Proposition 1 by a constant  in order to bring the principal part of these polynomials to standard form used in \cite{pw17a,grav04,pw17}.

Of course,  any polynomial combinations of $I_{1,2}$ and $X_N$ are also integrals of motion. For instance, there are other integrals of motion that are  functions on $(\omega_1-\omega_2)^2$. It is interesting that for $M_{1,2}\neq 0$  they are polynomials in momenta of less degree $N-2$.  For negative $M$ such integrals have the following form
\[
Y_{N-1}=
\dfrac{X_N^2}{I_1^{n_1}I_2^{n_2}}+\dfrac{\alpha I_1^{n_1}}{I^{n_2}}+\dfrac{\beta I_2^{n_1}}{I_1^{n_2}}=(\omega_1-\omega_2)^2I_1^{n_1}I_2^{n_2}+\dfrac{\alpha I_1^{n_1}}{I^{n_2}}+\dfrac{\beta I_2^{n_1}}{I_1^{n_2}}\,,
\]
where $\alpha,\beta$ are polynomials in $a,b$ and binomial coefficients.  For instance, Hamiltonian
\[
H= p_1^2+p_2^2+\dfrac{a}{q_1^2}+\dfrac{b}{q_2^{2/5}}
\]
is in the  involution with polynomial  in  momenta $X_{7}=(\omega_1-\omega_2) I_1I_2^ {3} $ of the seventh degree and with the following polynomial of the sixth degree
\[\begin{array}{rcl}
Y_6&=&4(\omega_1-\omega_2)^2I_1I_2^3+\dfrac{64b^5I_1}{9I_2^3}+\dfrac{aI_2^3}{I_1}\\ \\
&=&p_2^4(p_1q_2-p_2q_1)^2+\frac{a}{q_1^2}\left(q_2^2p_2^4
+\frac{11bq_2^{8/5}p_2^2}{3}+\frac{64b^2q_2^{6/5}}{9}\right)+\frac{b^3q_1^2}{q_2^{6/5}}\\
\\
&+&\frac{b^2(64p_1^2q_2^2-90p_1p_2q_1q_2+27p_2^2q_1^2)}{9q_2^{4/5}}
+\frac{bp_2^2(11p_1q_2-9p_2q_1)(p_1q_2-p_2q_1)}{3q_2^{2/5}}\,.
\end{array}
\]
More symmetric  Hamiltonian
\[
H=p_1^2+p_2^2+\dfrac{a}{q_1^{2/3}}+\dfrac{b}{q_2^{2/3}}
\]
is in involution with polynomial $X_7=(\omega_1-\omega_2) I_1^2I_2^ {2} $ of the seventh degree
and with the following  polynomial  of the sixth degree
\[\begin{array}{rcl}
Y_6&=&4I_1^2I_2^2(\omega_1-\omega_2)^2+\dfrac{4a^3I_2^2}{I_1^2}+\dfrac{4b^3I_1^2}{I_2^2}\\ \\
&=&p_1^2p_2^2(p_1q_2-p_2q_1)^2+2(p_1q_2-p_2q_1)\left( \dfrac{ap_2^2(q_2p_1-2q_1p_2)}{q_1^{2/3}}+\dfrac{bp_1^2(2q_2p_1-q_1p_2)}{q_2^{2/3}}\right)
\\
\\
&&+\dfrac{a^2q_2^2p_2^2}{q_1^{4/3}}
+\dfrac{2ab(4q_2^2p_1^2-9q_1q_2p_1p_2+4q_1^2p_2^2)}{q_1^{2/3}q_2^{2/3}}
+\dfrac{b^2q_1^2p_1^2}{q_2^{4/3}}+4ab\left(
\dfrac{aq_2^{4/3}}{q_1^{4/3}}+\dfrac{bq_1^{4/3}}{q_2^{4/3}}\right)\,.
\end{array}
\]
In a similar manner we can construct integrals of motion $Y_{N-1}$ for positive exponents $M$ and for composition of the positive and negative exponents.

\section{Nonseparable systems}
Let us start with the following theorem from  \cite{am16}.
\begin{prop}
If a Hamiltonian system defined by
\[
H=p_1^2+p_2^2+aq_1^{M_1}q_2^{M_2}
\]
is integrable in the Liouville sense, then either
\bq\label{am-cond}
M_1+M_2=\dfrac{2}{2p+1}\qquad\mbox{or}\qquad
M_1+M_2=\dfrac{2(p+1)}{p(p-1)}\,,\qquad p\in\mathbb{Z}
\eq
for a certain integer $p$.
\end{prop}
The conditions (\ref{am-cond})  are only necessary for the integrability. Only some of the potentials satisfying these conditions
are integrable.

We can obtain the known list of these integrable systems considering deformations of the Thompson integrals of motion
\bq\label{def-ham}
\tilde{H}=H+U(q_1,q_2)\,,\qquad \tilde{Z}_N=Z_N+\Delta Z_{N-2}\,,\qquad Z_N=F(I_1,I_2,\omega_1-\omega_2)\,.
\eq
Here $H$ is given by (\ref{ham-add}) at $M_1=0$, $Z_N$ is some fixed polynomial  in momenta of degree $N$, whereas potential $U(q_1,q_2)$ and polynomial  $\Delta Z_{N-2}$ of degree $N-2$ have to be obtained by solving equation
\[\{\tilde{H},\tilde{Z}_N\}=\sum_{i,j=1}^n\left( \dfrac{\partial \tilde{H} }{\partial q_i}\dfrac{\partial \tilde{Z}_N}{\partial p_j}- \dfrac{\partial \tilde{H}}{\partial p_j}\dfrac{\partial \tilde{Z}_N}{\partial q_i}\right)=0\,.
\]
For instance,  let us take
\bq\label{ham-23}
\begin{array}{rcl}
H&=&p_1^2+p_2^2+bq_2^{-2/3}\,,\qquad I_1=p_1^2\,,\qquad I_2=p_2^2+bq_2^{-2/3}\,,\\
\\
X_4&=&p_2^3(q_2p_1-q_1p_2) +p_2\left(3q_2p_1-2q_1p_2\right)bq_2^{-2/3}-b^2q_1q_2^{-4/3}
\\ \\
Y_4&=&p_2^2(q_2p_1-q_1p_2)^2 +(q_2p_1-q_1p_2)(2q_2p_1-4q_1p_2)bq_2^{-2/3}+b^2q_1^2q_2^{-4/3}\,,
\end{array}
\eq
where $X_4$ and $Y_4$ are integrals of motion considered in  previous Section.

It is easy to find deformation of this Hamiltonian
\bq\label{23-1}
\tilde{H}=p_1^2+p_2^2+\left(a q_1+b\right)q_2^{-2/3}\,,
\eq
which is in involution with  two functionally  independent integrals of motion
\[
\tilde{Z}_3=\sqrt{I_1}(3H-I_1)+\Delta Z_2=p_1(2p_1^2+3p_2^2)+\dfrac{3a(2p_1q_1+3p_2q_2)}{2q_2^{2/3}}+\dfrac{3 b p_1}{q_2^{2/3}}
\]
 and
\[
\tilde{Z}_4=I_1(2H-I_1)+\Delta Z_3=p_1^2(p_1^2+2p_2^2)+\dfrac{2ap_1(p_1q_1+3p_2q_2)}{q_2^{2/3}}+\dfrac{2b p_1^2}{q_2^{2/3}}+ \dfrac{9a^2q_2^{2/3}}{2}\,.
\]
Properties of this superintegrable system are discussed in \cite{ran13,pw11}.  Similar deformation
 \bq\label{1-23}
\tilde{H}=p_1^2+p_2^2+\left(a q_1^{-2/3}+b\right)q_2
\eq
is  integrable with first integral
\[
\tilde{Z}_4=I_1^4+\Delta Z_3=p_1^4+\dfrac{2ap_1(p_1q_2-3p_2q_1)}{q_1^{2/3}}-\dfrac{9a\beta q_1^{4/3}}{4}-\dfrac{a^2(9q_1^2-2q_2^2)}{2q_1^{4/3}}\,.
\]
On the one hand, both Hamiltonians (\ref{23-1}) and (\ref{1-23}) can be obtained  from the Hamiltonians of various  Holt systems  \cite{gram84,hiet87,holt82} using  shift  $q_1\to q_1+\alpha$. On the other hand, all the Holt systems can be considered as deformations (\ref{def-ham}) of the Thompson superintegrable system (\ref{ham-23}),  see \cite{ran13}.

Next integrable deformation  (\ref{def-ham})  of the same system (\ref{ham-23}) can be obtained using sixth order polynomial in momenta
\[
Z_6=I_1^2Y=p_1^2p_2^2(q_2p_1-q_1p_2)^2+\cdots
\]
which now depends on the angle variables. In this case solving equation  $\{\tilde{H},\tilde{Z}_N\}=0$ one gets a new integrable deformation of the Fokas-Lagerstrom system \cite{fl80}.

\begin{prop}
Hamiltonian
\bq\label{gen-fl}
\tilde{H}=p_1^2+p_2^2+\left(a q_1^{-2/3}+b\right)q_2^{-2/3}
\eq
is in involution with the following integral of motion
\[\begin{array}{rcl}
\tilde{Z}_6=Z_6+\Delta Z_{4}&=&p_1^2p_2^2(p_1q_2-p_2q_1)^2
-2a
\left(\frac{p_1p_2(p_1q_2-p_2q_1)(p_1q_1-p_2q_2)}{q_1^{2/3}q_2^{2/3}}
-\frac{ bp_1(p_1q_1^2+4p_1q_2^2-p_2q_1q_2)}{q_1^{2/3}q_2^{4/3}}\right)\\
\\
&+&a^2\left(\frac{(p_1 q_1-p_2 q_2)^2}{q_1^{4/3}q_2^{4/3}}+\frac{4 b}{q_1^{4/3}}\right) +\frac{2 b p_1^2(p_1q_2-p_2q_1)(2p_1q_2-p_2q_1)}{q_2^{2/3}}+\frac{b^2 q_1^2 p_1^2}{q_2^{4/3}}\,,
\end{array}
\]
which is polynomial in momenta of the sixth degree.
\end{prop}
For  $b=0$  potential in  $\tilde{H}$ (\ref{gen-fl}) coincides with the so-called Fokas-Lagerstrom  potential  \cite{fl80,hiet87}
\[
U=\dfrac{a}{(x^2-y^2)^{2/3}}\,,
\]
after a  45 degree rotation
\[
q_1=x-y\,,\qquad q_2=x+y\,.
\]
It is easy to directly prove that for other pairs of exponents $(M_1,M_2)$ in (\ref{ham-mul}), which satisfy conditions (\ref{am-cond}),  the first integral has to be a degree more than five in the momenta.

\section{Conclusion}
In this note we have carried out a systematic study of superintegrable Hamiltonian systems
separable in Cartesian coordinates using action-angle variables, which  play a fundamental role
in classical and quantum mechanics. It is enough to say that they are the key points in the Kolmogorov–Arnold–Moser theory,
 in the  geometric and  semi-classical quantization.

Previously in \cite{ts08,ts08a,ts09,ts12}, we have already constructed polynomial integrals of motion using addition theorems for the action-angle variables. For instance, by adding  action variables
\[
I_1=p_1^2+m^2q_1^2+aq_1\,,\qquad  I_2=p_2^2+n^2q_2^2+\dfrac{b}{q_2^2}\,,\qquad m,n,a,b\in\mathbb R\,,
\]
one gets Hamiltonian
\[
H=I_1+I_2=p_1^2+p_2^2+m^2q_1^2+n^2q_2^2+aq_1+\dfrac{b}{q_2^2}\,,
\]
which is in involution with the following  integral of motion
\[
X=F(I_1,I_2,\omega_2-\omega_1)\,,\qquad \{H,X\}=0\,,
\]
which is functionally independent from $I_{1,2}$. Here
\[
\omega_1= -\dfrac{1}{m}\arctan\left(\dfrac{2m^2q_1+a}{2mp_1}\right)\,,\qquad
\omega_2=\dfrac{1}{4n}\arctan\left(\dfrac{p_2^2-n^2q_2^2+bq_2^{-2}}{2nq_2p_2}\right)\,,
\]
are the corresponding action variables. For integer $m$ and half-integer $n$ this integral could be polynomial in momenta
\[\begin{array}{rcl}
X_{2n+m}&=&(-a^2-4m^2I_1)^n\, (4n^2b-I_2^2)^{m/2}\,e^{4\mathrm i m n(\omega_1-\omega_2)}\\
\\
&=&\left(2\,\mathrm{i}\,mp_1+2m^2q_1+a\right)^{2n}\left(2\,\mathrm{i}\, np_2q_2+p_2^2-n^2q_2^2+\dfrac{b}{q_2^2}\right)^m\,,\qquad
\mathrm{i}^2=-1\,,
\end{array}
\]
which is obtained using an addition theorem for logarithmic (inverse trigonometric) functions.

In this note we use the simplest addition theorems for polynomials (rational functions). Because we know how to add polynomials in quantum variables we could try to study  quantum counterpart of the Hamiltonian (\ref{ham-add} ) using quantum analogs of the action-angle variables \cite{sard05, lew96}. The main problem here is that  classical action-angle variables are defined only in some domain of the cotangent bundle of plane. In \cite{pw17a,nieto17, pw17} authors de facto found quantum  action-angle variables in the framework of the Bohr-Sommerfeld quantization in the Cartesian coordinates. It will be interesting to obtain these quantum  action-angle variable in geometric or semi-classical  quantization.

We are very grateful to the referees for  thorough analysis of the manuscript, constructive  suggestions and  proposed  corrections, which certainly lead to a more profound discussion of the results. The work was supported by the Russian Science Foundation (project 18-11-00032).

\end{document}